\documentclass[a4paper,11pt]{article}
\pdfoutput=1 
\usepackage{jcappub} 
\usepackage[T1]{fontenc} 
\usepackage{dcolumn}
\usepackage{float}
\usepackage{bm,tensor,braket} 
\usepackage{mathtools}
\usepackage{extarrows}
\usepackage[version=4]{mhchem}
\usepackage[utf8]{inputenc}
\usepackage{enumerate}
\usepackage{orcidlink}
\usepackage[capitalize]{cleveref}

\title{\boldmath Strangeon Ergostars}

\author[a]{Haojia Xia,}
\author[a,b]{Shichuan Chen,}
\author[b,1]{Hong-Bo Li,\note{Corresponding author.}}
\author[c,1]{Enping Zhou,}
\author[d,a,b,1]{and Ren-Xin Xu}

\affiliation[a]{Department of Astronomy, School of Physics, Peking University, Beijing 100871, China}
\affiliation[b]{Kavli Institute for Astronomy and Astrophysics, Peking University, Beijing 100871, China}
\affiliation[c]{School of Physics, Huazhong University of Science and Technology, Wuhan 430074, China}
\affiliation[d]{State Key Laboratory of Nuclear Physics and Technology, Peking University, Beijing 100871, China}

\emailAdd{lihb2020@pku.edu.cn}
\emailAdd{ezhou@hust.edu.cn}
\emailAdd{r.x.xu@pku.edu.cn}

\abstract{The nature of the central engine powering short gamma-ray bursts (sGRBs) in binary neutron star (BNS) mergers remains a key open question in the era of multi-messenger astronomy.
The ergostar hypothesis, that a rapidly rotating star with an ergoregion can act as a powerful energy source, offers an alternative explanation to the black hole-accretion disk paradigm. In this work, however, we examine this hypothesis using a phenomenological EOS of strangeon matter, i.e., condensed matter with nucleon-like units for three flavors of quarks. By constructing a large suite of uniformly rotating equilibrium models, we systematically investigate the parameter space of the stable ergostars and calculate their maximum extractable energy. We demonstrate that strangeon matter supports a vast and robust parameter space for dynamically stable ergostars, even without requiring differential rotation. We find that the extractable rotational energy from these configurations can be on the order of $0.01 M_\odot$, representing a massive energy reservoir, even when accounting for baryonic mass variations (e.g., mass ejection and particle capture during the Penrose process).  Our results suggest that BNS merger remnants composed of exotic matter could play a crucial, previously underestimated role in high-energy astrophysics.}

\begin{document}
\maketitle
\flushbottom

\section{Introduction}
\label{sec: intro}

Short gamma-ray bursts (sGRB) are among the most violent astrophysical phenomena in the universe.
The multi-messenger observation of the binary neutron star (BNS) merger GW170817 provided decisive evidence that such mergers are a primary progenitor of sGRB \cite{GW170817, Savchenko_2017, 2017GCN}. In the standard paradigm, the central engine is a nascent black hole surrounded by an accretion disk \cite{Blandford1977}.
However, the viability of alternative scenarios—particularly those involving a long-lived, hypermassive neutron star (or magnetar) remnant—remains an open and important question \cite{Dai_1998, Zhang_2001, Zhang_2015}.

Alternatively, Komissarov~\cite{Komissarov2004, Komissarov2005} proposed that the central engine could be a rapidly rotating compact star containing an ergoregion, often referred to as an ergostar. In general relativity, the ergoregion is a region of spacetime where the stellar rotation is so strong that no particle can remain stationary with respect to a distant observer. Instead, all trajectories, including those of light, are forced to co-rotate with the star. Mathematically, this corresponds to the condition where the metric component $g_{\rm tt} > 0$.

Such extreme frame-dragging allows for the existence of negative-energy states relative to infinity.
These states open a physical channel for energy extraction from a BNS merger remnant, analogous to the Penrose process for black holes \cite{Penrose2002}, provided that the ergostar remains stable for a sufficient duration. If the configuration proves unstable, the ergoregion may disappear, or the star may collapse entirely, before substantial energy extraction occurs.

This stability problem is intrinsically connected to the equation of state (EOS) of dense matter. 
While ergostar models have been discussed for decades, such as the seminal differentially rotating polytropic models by Komatsu, Eriguchi, and Hachisu (KEH) \mbox{\cite{KEH1, KEH2}}, recent full general relativistic simulations by Tsokaros \emph{et al.}~\mbox{\cite{Tsokaros2019}} demonstrated that these KEH ergostars are unfortunately dynamically unstable. Furthermore, for conventional nuclear matter EOSs (where the density drops to zero at the surface), dynamically stable ergostars are extremely rare and typically require substantial differential rotation \mbox{\cite{Tsokaros2020b}}. However, Tsokaros \emph{et al.}~\mbox{\cite{Tsokaros2019, Tsokaros2020b}} also revealed a crucial mechanism: the most important factor for the existence of a stable ergosphere is having a sufficiently large density close to the surface of the compact star, resembling the density profile of a bare quark star \mbox{\cite{Tsokaros2019, Tsokaros2020b, Tsokaros2020c, Tsokaros2020a}}. When this condition is met, differential rotation becomes secondary, and stable ergostars can readily form.

In fact, the true EOS of supranuclear matter remains still unknown due to the difficulty from ab initio calculations. Nevertheless, strangeness matters~\cite[e.g.,][]{2023AdPhX...837433L, 2025IJMPA..4050180X}.
Witten~\cite{Witten:1984rs} proposed that the true ground state of dense matter might be quark matter, composed of nearly free $u$, $d$, and $s$ quarks. This idea implies that compact objects resembling pulsars could be strange quark stars (SqSs) rather than traditional neutron stars (NSs), as modeled in the MIT bag model within the asymptotic freedom regime \cite{Alcock:1986hz}.
Can quarks really be asymptotically free there?
Could nucleon-like units form due to the non-perturbative nature of the fundamental strong interaction at low energies as in the case of atomic nucleus?
Xu~\cite{Xu:2003xe} suggested then that the building blocks of supranuclear matter could instead be strange quark clusters, termed strangeons,\footnote{
This kind of nucleon-like bound state could also have a dramatic impact on our understanding of the material world in the present universe. See~\cite{2025Univ...11..354Q} for a brief review.
} %
rather than itinerant quarks. This is because the coupling between quarks remains very strong \cite{Lai:2017ney}. These strangeons, formed by bound $u$, $d$, and $s$ quarks, represent a unique state of matter in which quarks condense in position space rather than momentum space. The term strangeon stars (SnSs) was coined to describe such compact objects \cite{Xu:2016uod, Lai:2017ney, 2022MNRAS.516.6172L, Yuan2025}.

Strangeon matter, like strange quark matter, consists of nearly equal numbers of $u$, $d$, and $s$ quarks. In contrast to strange quark matter, quarks in strangeon matter are localized inside strangeons due to the strong quark-quark coupling. NSs, SqSs and SnSs have similarities but also differ significantly. In SnSs, quarks are localized in strangeons, much like neutrons in NSs. However, unlike neutrons, a strangeon can contain more than three valence quarks, restoring light-flavour symmetry. SnSs are self-bound by the strong force, with their surface matter also composed of strangeons, similar to SqSs \cite{Xu:2003xe}.

The SnSs model can account for many key observational phenomena in astrophysics. The EOS of SnSs is sufficiently stiff to explain the observed masses of pulsars \cite{Demorest:2010bx, Antoniadis:2013pzd}, while pulsar glitches may be naturally attributed to starquakes \cite{Zhou:2014tba}. Moreover, SnSs provide a potential explanation for X-ray flares and bursts in magnetar candidates \cite{Xu:2006mp}, the plateau phase observed in gamma-ray bursts \cite{2011SCPMA..54.1541D}, and the quasi-periodic oscillations in SGR 1806$-$20 \cite{Li:2023tng}.
They also help to shape the energy budget of various astrophysical power sources within the framework of magnetar starquake triggering mechanisms \cite{Wang:2024opz}.
In addition, the strangeness barrier plays a crucial role in understanding Type I X-ray bursters \cite{2015ApJ...798...56L}, X-ray-dim isolated neutron stars \cite{Wang:2016nqt, Wang:2017hgt}, and the nature of the Ultraluminous X-ray sources \cite{2026MNRAS.546ag241L}.

In this work, we investigate the dynamic stability of the ergostars based on the SnS model. We utilize the publicly available \texttt{rns} code to construct equilibrium sequences of rotating stars and identify ergostar solutions using the $g_{tt}>0$ criterion. The dynamical stability of these solutions is analyzed via the turning-point method~\cite{Friedman1988}. On this basis, we estimate the maximum extractable rotational energy by calculating the amount of energy loss required for the ergoregion to vanish or for the star to collapse into a black hole. We find that strangeon matter supports a robust population of dynamically stable ergostars with a massive available energy budget ($\sim 10^{52}\, \rm erg$). 

The paper is organized as follows. In section~\ref{sec:EOS}, we introduce the EOS of SnSs adopted in this work.
We discuss in section~\ref{sec: Results} the results of our systematic survey, identifying the parameter space for stable ergostars and calculating their maximum extractable energy.
Finally, we summarize and discuss our results in section~\ref{sec: Discussion}.

\section{Equation of state}
\label{sec:EOS}

As an analogy of the interaction between nucleons, we assume that the interaction potential between two strangeons is parameterized by the Lennard-Jones potential~\cite{Xu2003, Lai2009, Gao2021, Li:2022qql, Zhang2023_9, Zhang2023_12, Yuan2025},
\begin{equation}
    U(r) = 4\epsilon \left[ \left(\frac{\sigma}{r}\right)^{12} - \left(\frac{\sigma}{r}\right)^{6} \right] \,,
\end{equation}
where $r$ is the distance between two strangeons, $\epsilon$ is the depth of the potential well representing the interaction strength, and $\sigma$ is the characteristic distance scale where the potential is zero.

From this two-body potential, the energy density $\rho$ and pressure $P$ of bulk strangeon matter at zero temperature can be derived by assuming a crystal lattice structure\cite{Lai2009}:
\begin{equation}
\begin{aligned}
& \rho=2 \epsilon\left(A_{12} \sigma^{12} n^5-A_6 \sigma^6 n^3\right)+n N_{\mathrm{q}} m_{\mathrm{q}} \,, \\
& P=n^2 \frac{\mathrm{d}(\rho / n)}{\mathrm{d} n}=4 \epsilon\left(2 A_{12} \sigma^{12} n^5-A_6 \sigma^6 n^3\right) \,, \label{eq:eos_three}
\end{aligned}
\end{equation}
where $A_{12}=6.2, A_6=8.4$, $n$ is number density of strangeons, $N_q$ is number of quarks in a strangeon, and $m_q$ is the average constituent quark mass. 

The EOS can be recast in terms of two parameters through a normalization procedure. This procedure begins by replacing the microscopic parameter $\sigma$ with the macroscopic surface baryon number density, $n_{\text{sur}}=\left(A_6/2 A_{12}\right)^{1 / 2}N_{\rm q}/3\sigma^3$, the density at which the pressure vanishes. Then, the potential depth $\epsilon$ and the number density $n$ are normalized. The normalized potential depth is defined as $\tilde{\epsilon} = \epsilon / N_q$, representing the potential depth contribution per quark, which serves as a measure of interaction strength independent of the strangeon's size. The number density is normalized as $\bar{n}$, relating the local density to the surface density.

This process yields the final ``two-parameter model'' form used in our analysis \cite{Zhang2023_12}:
\begin{align}
    \frac{\rho}{n_{\text{sur}}} &= \frac{a}{9}\tilde{\epsilon}\left(\frac{1}{18}\bar{n}^5 - \bar{n}^3\right) + m_q\bar{n} \,, \label{eq:eos_rho} \\
    \frac{P}{n_{\text{sur}}} &= \frac{2a}{9}\tilde{\epsilon}\left(\frac{1}{9}\bar{n}^5 - \bar{n}^3\right) \,.
\label{eq:eos_p}
\end{align}
Here, the normalized number density is $\bar{n} = (N_q n) / n_{\text{sur}}$. The model contains several fixed constants. Following the original paper\cite{Yuan2025,Zhang2023_9}, the number of quarks per strangeon, $N_q = 18$, the parameter $a$ derived from lattice sum constants of the Lennard-Jones potential, is defined as $a = A_6^2 / A_{12} = 8.4^2 / 6.2 \approx 11.38$, the average constituent quark mass $m_q$ is taken to be $300\, \rm MeV$.

The entire thermodynamic behavior of the strangeon matter is uniquely determined by two free parameters. The first is the normalized potential depth, $\tilde{\epsilon}$ (in MeV), which primarily controls the stiffness of the EOS, a larger $\tilde{\epsilon}$ corresponds to stronger short-range repulsion and thus a stiffer EOS. The second parameter is the surface baryon number density, $n_{\text{sur}}$ (in fm$^{-3}$), which sets the saturation density of the self-bound matter. A lower value of $n_{\text{sur}}$ also contributes to a stiffer EOS.

An important feature of this EOS, inherited from its non-relativistic form of Lennard–Jones potential, is that the so-called adiabatic speed of sound, defined as $c_s^2 = dP/d\rho$, can exceed the speed of light at high densities. This seemingly acausal behavior is a known characteristic of models employing classical, action-at-a-distance potentials \cite{Bludman1968, Caporaso1979}. Following the established arguments \cite{Caporaso1979, Lai2009}, it is important to distinguish this adiabatic sound speed from the true physical signal propagation speed. The quantity $c_s$ is derived from the static equilibrium relation of the EOS and should therefore be interpreted as a measurement of the stiffness of the matter rather than the propagation speed of a physical disturbance.

In phenomenological models based on classical interaction potentials, such as the Lennard-Jones-type interaction adopted here, the interparticle forces are effectively treated as instantaneous. As discussed by  Bludman and Ruderman~\cite{Bludman1968} and Caporaso and Breche~\cite{Caporaso1979}, this approximation leads to a thermodynamic derivative $dP/d\rho$ that formally exceeds $c^2$. However, this does not imply a violation of relativistic causality. More importantly, for the specific case of strangeon matter, Lu \emph{et al.}~\cite{Lu2018} have explicitly calculated the actual physical signal speed by analyzing the time-domain impulse response in a discrete lattice model with retarding interactions. They rigorously demonstrated that the true signal propagation speed $c_{\rm signal}$ is bounded by
\begin{equation}
c_{\rm signal} \approx \left( \frac{1}{\sqrt{(\partial P/\partial \rho)_S}} + \frac{1}{c} \right)^{-1} < c \,.
\end{equation}
Thus, the actual causal signal speed remains subluminal regardless of the macroscopic derivative $dP/d\rho$.

An order-of-magnitude estimation shows that the strangeons inside condensed matter tend to be particle-like. This particle-like property could be quantified by the ratio of the characteristic kinetic energy, $E_{\rm k}$, to the potential depth, $U$.
For a particle with mass $m$, confined within a separation $r$, one has $E_{\rm k} \sim \hbar^2/(2 m r^2)$, and then
\begin{equation}
\delta \equiv \frac{U}{E_{\rm k}}  \sim \frac{2m r^2 U}{\hbar^2} \sim 0.5m_{\rm GeV}U_{\rm 10}r_1^2 \,,
\end{equation}
where $m=m_{\rm GeV}$\,GeV, $U=10U_{\rm 10}$\,MeV and $r=r_{\rm 1}$\,fm.
This suggests that strangeon matter, compared to nucleon matter, could be better described by a potential force acting on classical particles,\footnote{%
This results in the conjecture of classical solid (rather than super-solid~\cite{2007PhRvD..76g4026M}) for cold strangeon matter~\cite{Xu:2003xe}, though historically, the atomic nucleus was also speculated in a solid state~\cite{1974AnPhy..86..138B}.
} %
and consequently, we have large $c_s$ since $m_{\rm GeV}\sim 1$ for nucleon whereas $m_{\rm GeV}\gtrsim 10$ for strangeon~\cite{Yuan2025}.

Throughout this work, we therefore interpret $c_s$ as a thermodynamic quantity characterizing the stiffness of the EOS. The apparent superluminal value of $c_s$ arises from the idealized instantaneous interaction assumed in the effective potential model, while the fundamental causality of the underlying strong interaction is strictly preserved as explicitly demonstrated in previous works \cite{Lu2018}. It is worth noting that macroscopic numerical relativity (NR) simulations cannot be used to calculate or verify this true physical signal speed. In any NR framework, the EOS and its corresponding sound speed are required as inputs to close the general relativistic hydrodynamics equations. An NR simulation merely evolves the system according to these provided macroscopic inputs, rather than determining the fundamental causality limit, which must be derived from microscopic many-body physics. Consequently, for our main systematic survey of uniformly rotating stars, we do not artificially restrict the parameter space by enforcing a mathematical $c_s \le c$ limit. Nevertheless, to ensure the absolute robustness of our proposed astrophysical mechanism and to dispel any potential concerns regarding relativistic causality, we will first explicitly verify our central engine model against a stringent test case that strictly imposes $c_s \le c$ in section~\ref{sec: Results}.

To systematically investigate the properties of this EOS and the potential for ergostar formation, we explore a broad parameter space, extending beyond the specific regions constrained by recent Bayesian analyses \cite{Yuan2025}. We select a representative series of values for the normalized potential depth, $\tilde{\epsilon}$: 0.1, 1.0, and 3.0 MeV. These values correspond to scenarios that are, respectively, softer, slightly stiffer, and significantly stiffer than the observationally favored results. For each fixed value of $\tilde{\epsilon}$, we generate a sequence of EOSs by uniformly sampling 30 points for the surface baryon number density, $n_{\text{sur}}$, within the range of $(0.1, 0.36)~\text{fm}^{-3}$. This comprehensive grid allows us to robustly map the scaling relations, identify the dynamically stable domain, and evaluate the energy extraction potential of strangeon ergostars.

\section{Results}
\label{sec: Results}

\subsection{Overview of Ergostar Analysis}

To investigate the properties of ergostars, we perform a systematic analysis for the EOS model.
Our equilibrium models of rotating compact stars are constructed using the publicly available \texttt{rns} code \cite{Stergioulas_1995}. This code solves the Einstein field equations for a stationary and axisymmetric spacetime, whose metric is given by
\begin{equation}
\begin{split}
    {\rm d} s^2=-e^{\gamma+ \beta} {\rm d} t^2 + e^{2 \alpha} \bigl({\rm d} r^2 + r^2 {\rm d} \theta^2 \bigr) + e^{\gamma - \beta} r^2 \sin^2 \theta \bigl( {\rm d} \phi - \omega {\rm d} t \bigr)^2 \,,
\end{split}
\end{equation}
where $\gamma$, $\beta$, $\alpha$, and the frame dragging frequency $\omega$ are functions of the coordinates $r$ and $\theta$. The energy-momentum tensor is given by
\begin{equation}
    T^{\mu\nu} = (\rho + P) u^\mu u^\nu + P g^{\mu\nu} \,,
\end{equation}
where $P$ is the pressure and $\rho$ is the energy density. In this work, we employ the phenomenological strangeon EOS, the details and parameterizations of which are presented in section~\ref{sec:EOS}.

For the majority of this work, we assume that the stars are in rigid (uniform) rotation, which means the angular velocity $\Omega$ is constant throughout the star. The computational procedure is as follows: for a given EOS, we generate a sequence of equilibrium solutions by incrementally increasing the central baryon density, $\rho_c$. For each resulting stellar model, we then test for the existence of an ergoregion by evaluating the sign of $g_{tt}$ across the stellar interior. An ergoregion is identified where the following condition is satisfied:
\begin{equation}
\label{eq:ergo}
    g_{tt} = -e^{\gamma+\beta} + e^{\gamma-\beta} r^2 \sin^2\theta \, \omega^2 > 0 \,.
\end{equation}

For each EOS, the analysis proceeds via a two-dimensional parameter space scan over the angular momentum $J$ and central baryon density $\rho_c$. We map the resulting equilibrium solutions onto the gravitational mass--central density ($M$--$\rho_c$) plane. In this plane, the domain of dynamically stable ergostars is delineated by the intersection of two physical conditions. First, the solution must satisfy the condition $g_{tt} > 0$ (eq.~\ref{eq:ergo}) to possess an ergoregion. Second, the solution must lie on the stable branch of the equilibrium sequence, defined by the turning-point criterion \cite{Friedman1988}. Based on this domain, we isolate the stable ergostar population and identify two critical boundary solutions: the stable ergostar with the minimum gravitational mass ($M_{\text{min-ergo}}$) and the one with the maximum gravitational mass ($M_{\text{max-ergo}}$).

After projecting these stable solutions onto the angular momentum--mass ($J$--$M$) plane, the stable ergostar solutions are strictly confined between the points corresponding to $M_{\text{min-ergo}}$ and $M_{\text{max-ergo}}$. This feature provides a practical diagnostic for astrophysical viability. By overlaying this stable ergostar region with the specific $J$--$M$ relation imposed by the BNS merger process—specifically the angular momentum budget of the remnant—one can rapidly assess whether a specific EOS is capable of supporting stable ergostars consistent with post-merger conditions.

\subsection{Extractable Energy}
\label{sec:energy}

To connect these theoretical models to the astrophysical outcomes of BNS mergers, we employ the empirical relation derived by Bauswein and Stergioulas~\cite{Bauswein_2017}. The empirical relation between angular momentum of the remnant and the total binary mass is largely insensitive to the underlying EOSs:
\begin{equation}\label{eq:jmerger-mtotal}
J_\mathrm{merger} \simeq a \,M_\mathrm{tot} - b \,,
\end{equation}
where $J_\mathrm{merger}$ is the angular momentum of merger remnant, $M_\mathrm{tot}$ is total binary mass, and the fitting constants are $a=4.041$ and $b=4.658$. To apply this relation, we adopt the approximation that the gravitational mass of the remnant is equal to the total binary mass ($M_\mathrm{remnant} \approx M_\mathrm{tot}$), neglecting the small percentage of mass lost during the merger process.

It should be noted that eq.~(\ref{eq:jmerger-mtotal}) was originally derived from numerical simulations of conventional neutron stars. However, full general relativistic hydrodynamic simulations of self-bound bare quark stars \cite{Zhou2022} have explicitly demonstrated that the angular momentum contained in the merger remnant is highly consistent with this empirical relation. As further corroborated by effective-one-body analyses \cite{Zhou2024}, the inspiral dynamics and the resulting angular momentum budget are predominantly governed by macroscopic properties (e.g., total mass and tidal deformability) rather than the surface density discontinuity. Given that the relation between the angular momentum acquired by the merger remnant and the total binary mass is predominantly governed by the dynamics of this inspiral phase, eq.~(\ref{eq:jmerger-mtotal}) remains a robust and accurate approximation for strangeon stars despite their non-zero surface density.

Building on this scenario, we hypothesize an energy-extraction process that proceeds along a sequence of constant baryon mass. Each evolutionary path begins from a configuration located below the merger-remnant line and terminates at the point of minimum gravitational mass along that sequence, provided it remains within the stable ergostar region. The released energy for a given path is then evaluated as
\begin{equation}
\Delta E \equiv \Delta(M - M_0) = (M_{\text{initial}} - M_{0,\text{initial}}) - (M_{\text{final}} - M_{0,\text{final}}) \,.
\end{equation}

While the above expression reduces to the gravitational mass difference for constant baryon mass sequences, the actual post-merger stage often involves changes in baryon mass (e.g., mass ejection). The physical rationale for this formulation becomes clear when considering the energetics of mass ejection. Although the total energy carried away by the ejecta is not exactly equal to its baryonic rest mass $M_{\rm ej0}$, it is strongly dominated by it. In reality, the ejecta also carries kinetic, internal, and gravitational potential energies. However, the asymptotic kinetic energy of typical dynamical ejecta in BNS mergers is only a small fraction of its rest mass ($E_k \sim 0.05 M_{\rm ej0} c^2$). The internal energy of the expanding, decompressing ejecta is similarly subdominant. Consequently, $\Delta(M-M_0)$ is a reliable metric for the rotational energy release.

This metric enables us to identify the trajectory that maximizes the energy release and to quantify its capacity. Crucially, the ergoregion acts as an independent, geometrically originated energy budget. While conventional mass ejection inevitably occurs during the merger, the rotational energy stored within the ergoregion constitutes an additional energy channel. 

While our fiducial estimation calculates the extraction along an idealized constant baryon mass sequence ($M_0 = \mathrm{const}$), a realistic BNS merger environment involves a complex interplay of mass-loss and mass-gain mechanisms. On one hand, mass ejection (e.g., dynamical ejecta and disk winds) reduces the remnant's baryon mass. On the other hand, the energy extraction mechanism itself---analogous to the Penrose process---typically increases the star's baryon mass. Specifically, the fragments that fall into the star from the ergoregion carry negative conserved energy (reducing the star's total gravitational mass $M$) but still possess a positive rest mass. Therefore, the capture of these negative-energy particles, along with potential fallback accretion, adds to the total baryon mass ($M_0$) of the remnant.

To accommodate these competing physical realities and provide a conservative estimate, we evaluate the energy extraction bounds by allowing the remnant's rest mass to fluctuate by $\Delta M_0 = \pm 0.01 M_\odot$ during the ergoregion phase. This bidirectional mass fluctuation introduces a bounded variation in the final extractable energy. As we will demonstrate in the following models, even when accounting for this $\pm 0.01 M_\odot$ fluctuation, the evolutionary trajectory of the ergostar remnant remains largely within the stable ergostar parameter space, ensuring that a substantial fraction of this energy reservoir remains viable.

Applying this energy-extraction framework, we first construct a stringent test model to address potential concerns regarding relativistic causality upfront. In this setup, we impose a strict relativistic causality correction (capping the adiabatic sound speed at $c_s \le c$) and incorporate differential rotation (with a degree of differential rotation $A=5$). The differential rotation is described by the Komatsu-Eriguchi-Hachisu law \cite{KEH1, KEH2} $ u^t u_\phi = A^2 (\Omega_c - \Omega) $, where $A$ is a parameter that controls the degree of differential rotation, and $\Omega_c$ is the angular velocity at the center of the star.

\begin{figure}
    \centering
    \includegraphics[width=0.9\columnwidth]{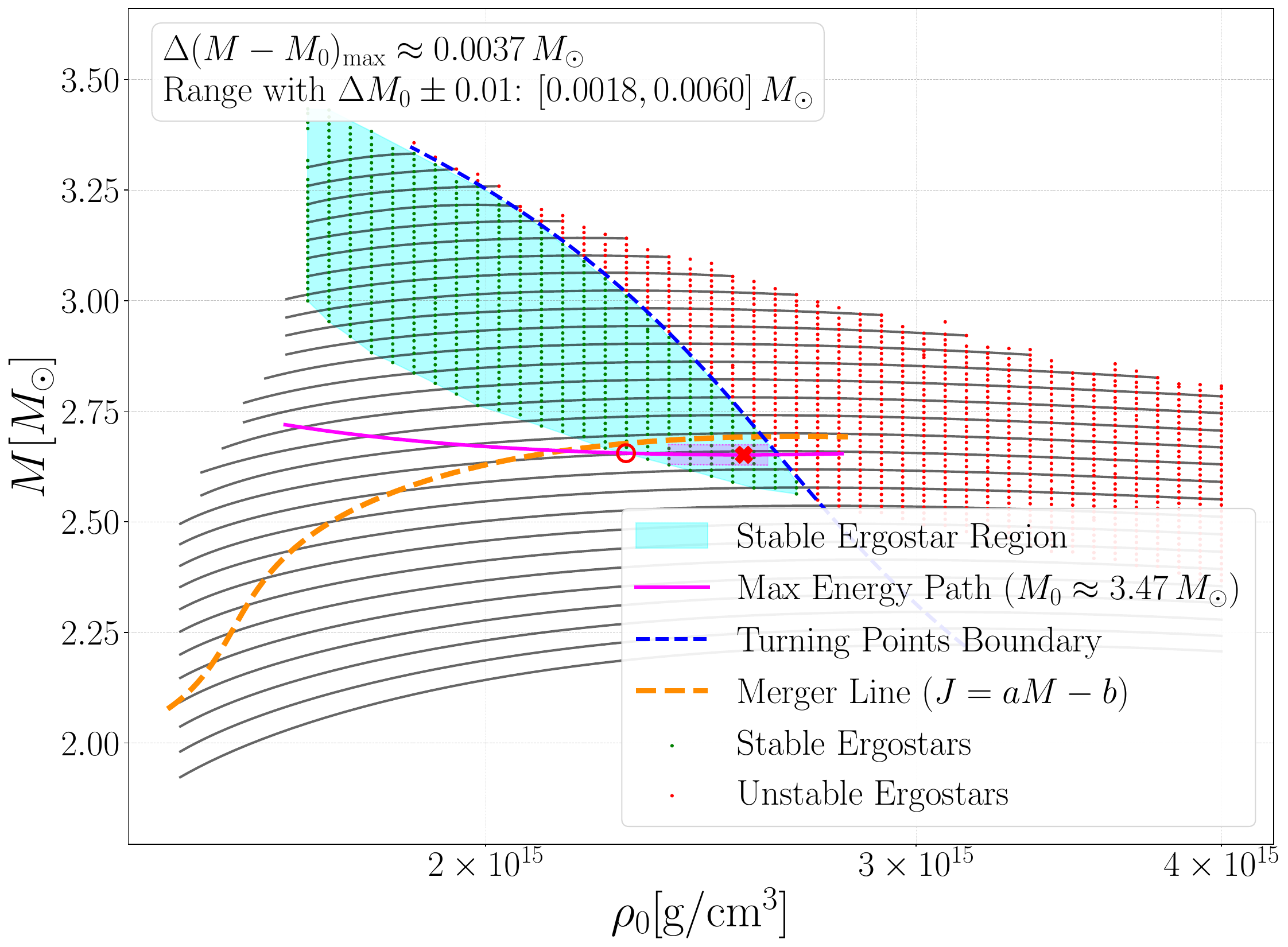}
    \caption{Ergostar solutions for a supplementary model with strict causality correction ($c_s \le c$) and differential rotation ($A=5$), using EOS parameters $\tilde{\epsilon}=0.1 \, \rm MeV$ and $n_{\text{sur}}=0.36\, \rm fm^{-3}$. The solid magenta line represents the fiducial maximum energy extraction path, while the shaded band illustrates its allowed variation range incorporating a baryonic mass fluctuation of $\Delta M_0 = \pm 0.01 M_\odot$ (accounting for mass ejection and Penrose particle capture). The red open circle and the red cross represent the starting and ending points, respectively, of the fiducial maximum energy extraction path without considering baryonic mass fluctuations. The persistence of the stable ergostar region indicates that our main conclusions are robust against causality constraints.}
    \label{fig:causality_corrected}
\end{figure}

As shown in figure~\ref{fig:causality_corrected}, even under these strict relativistic constraints and with the inclusion of the aforementioned baryonic mass fluctuations ($\Delta M_0 = \pm 0.01 M_\odot$), the formation of dynamically stable ergostars is completely preserved. The parameter space for stable ergostars persists, yielding a comparable maximum extractable energy ($\Delta M_{\rm max} \approx 0.0037 M_\odot$). This confirms that the existence of post-merger ergostars in our framework is a genuine consequence of the extreme stiffness of strangeon matter.

However, as extensively discussed in section~\ref{sec:EOS}, the adiabatic sound speed $c_s = \sqrt{dP/d\rho}$ derived from the phenomenological Lennard-Jones potential is merely a thermodynamic measure of the EOS stiffness rather than the true physical signal propagation speed. The actual physical signal speed remains strictly subluminal due to the retarding nature of the interactions \cite{Lu2018}. Therefore, the artificial mathematical constraint $c_s \le c$ can be safely relaxed.

Taking advantage of the extreme stiffness of strangeon matter, we find that if we do not artificially restrict the adiabatic sound speed, dynamically stable ergostar can form and provide the required magnitude of energy release \textit{even under the strict assumption of uniform rotation}.

\begin{figure}
    \centering
    \includegraphics[width=0.9\columnwidth]{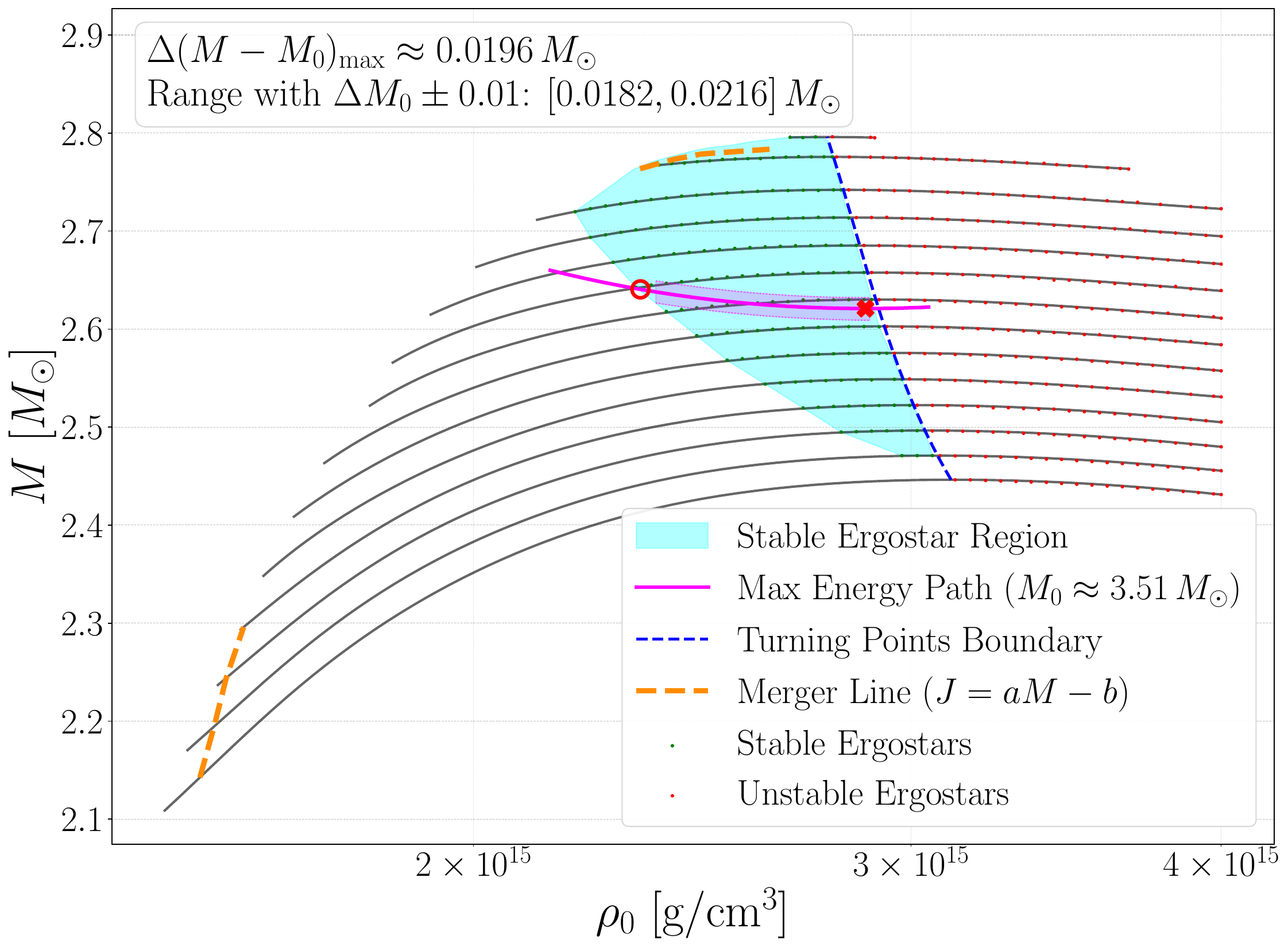}
    \caption{Gravitational mass versus central density for a uniformly rotating strangeon star with EOS parameters $\tilde{\epsilon}=0.1$~MeV and $n_{\text{sur}}=0.36\, \rm fm^{-3}$ (without limiting the adiabatic sound speed). Sequences of constant angular momentum are shown as black solid lines, with their turning points connected by the blue dashed line. The cyan region highlights the inferred domain of stable ergostars. The orange dashed line is the empirical relation for BNS merger remnants. The gold star marks the minimum-mass dynamically stable ergostar that can be formed from a BNS merger for this EOS, with its properties listed in the top-left corner. The solid magenta line indicates the path of maximum energy extraction, with the red open circle and red cross denoting its starting and ending points, respectively, as defined in figure~\ref{fig:causality_corrected}.}
    \label{fig:0.1_0.36}
\end{figure}

To illustrate this, we select a highly representative uniformly rotating model from our parameter space, characterized by $\tilde{\epsilon}=0.1\, \rm MeV$ and $n_{\text{sur}}=0.36\, \rm fm^{-3}$. As shown in figure~\ref{fig:0.1_0.36}, the stable ergostar domain spans a broad region. 

Furthermore, it is important to address the physical uniqueness of the strangeon ergostar model. As pointed out by Tsokaros \emph{et al.}~\mbox{\cite{Tsokaros2020b}}, phenomenological EOSs (such as the SLycc1 model) can also macroscopically produce stable ergostars with similar extractable energies by artificially imposing a finite surface density. While such models serve as excellent mathematical proofs-of-concept for the general relativistic mechanism, exploring physically originated EOSs is crucial to understanding whether the required mass and angular momentum can be realistically achieved in binary mergers. The strangeon EOS, derived from the non-perturbative nature of the strong interaction, naturally provides a vast and robust parameter space that is highly compatible with the post-merger remnant. More importantly, possessing a self-consistent physical model is essential for predicting other unique multi-messenger signatures, which will be the ultimate key to observationally identifying the true nature of compact stars. 

\subsection{Scaling Relations of Ergostar Solutions}
\label{sec:scaling}

\begin{figure}
    \centering
    \includegraphics[width=0.85\columnwidth]{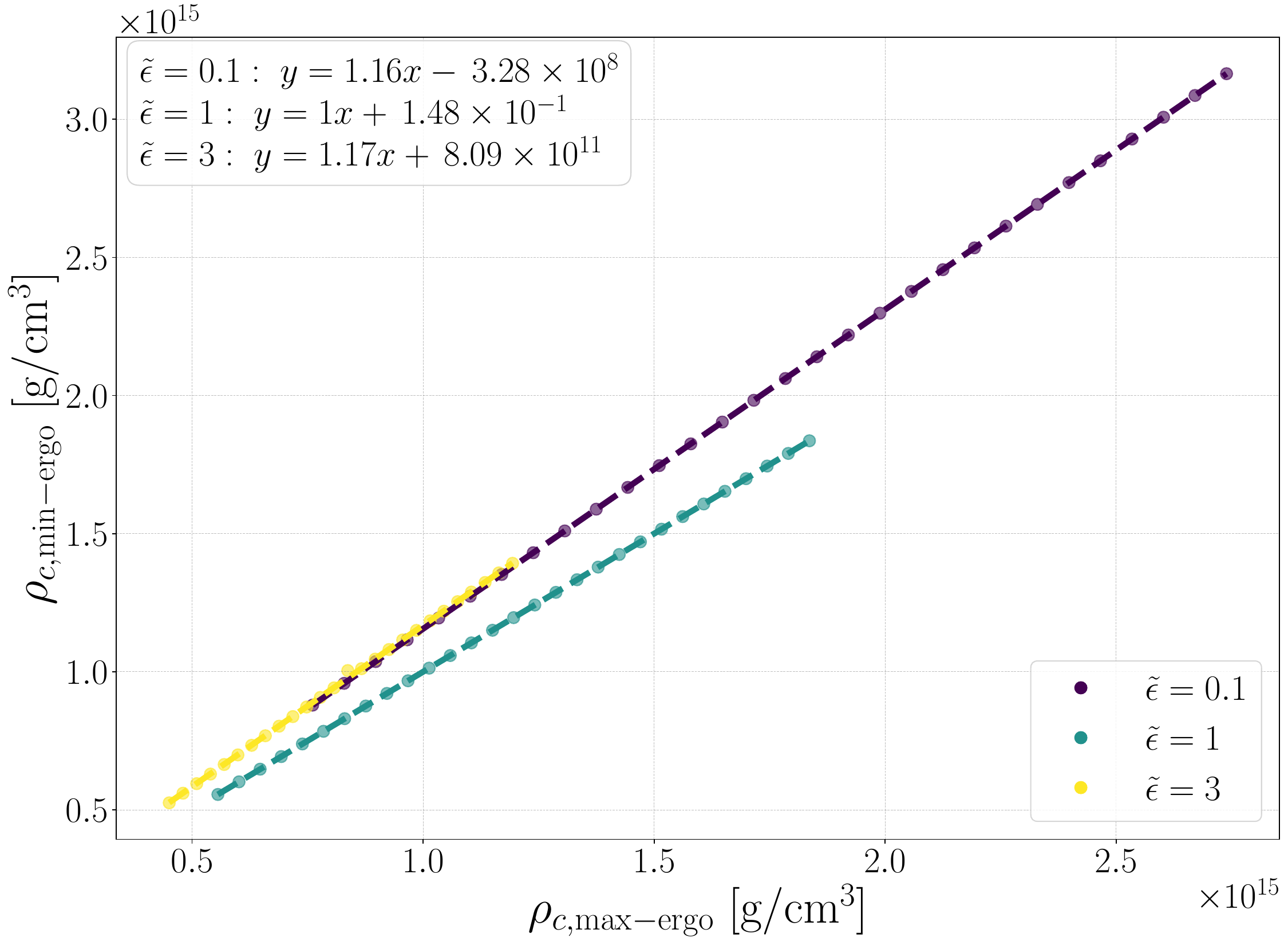}
    \caption{Scaling relation between the central density of the minimum-mass stable ergostar ($\rho_{c,\text{min-ergo}}$) and the central density of the maximum-mass stable ergostar ($\rho_{c,\text{max-ergo}}$) for each EOS parameter set. Each color corresponds to a different fixed value of the potential depth per quark, $\tilde{\epsilon}$. The solid points denote individual parameter combinations from our systematic survey, while the dashed lines represent the linear fits for each family of models. The best-fit formulas are displayed in the plot.}
    \label{fig: rho_rho}
\end{figure}

\begin{figure}
    \centering
    \includegraphics[width=0.85\columnwidth]{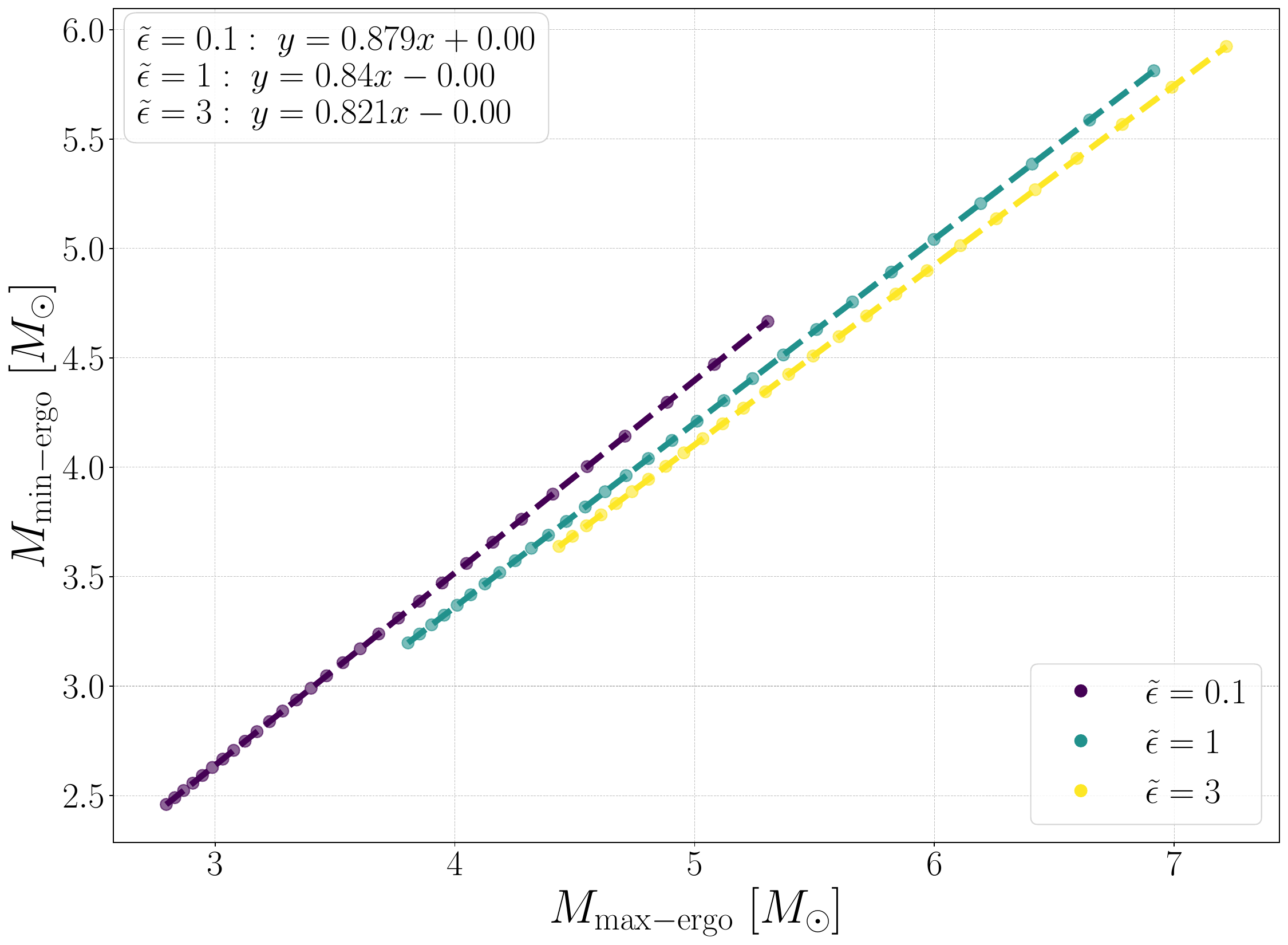}
    \caption{Scaling relation between the gravitational mass of the minimum-mass stable ergostar ($M_{\text{min-ergo}}$) and the gravitational mass of the maximum-mass stable ergostar ($M_{\text{max-ergo}}$). As in the previous figure, each color corresponds to a fixed value of $\tilde{\epsilon}$. The solid points represent individual parameter combinations, and the dashed lines are their respective linear fits. The fitting formulas are shown in the plot.} 
    \label{fig: m_m}
\end{figure}

\begin{figure}
    \centering
    \includegraphics[width=0.85\columnwidth]{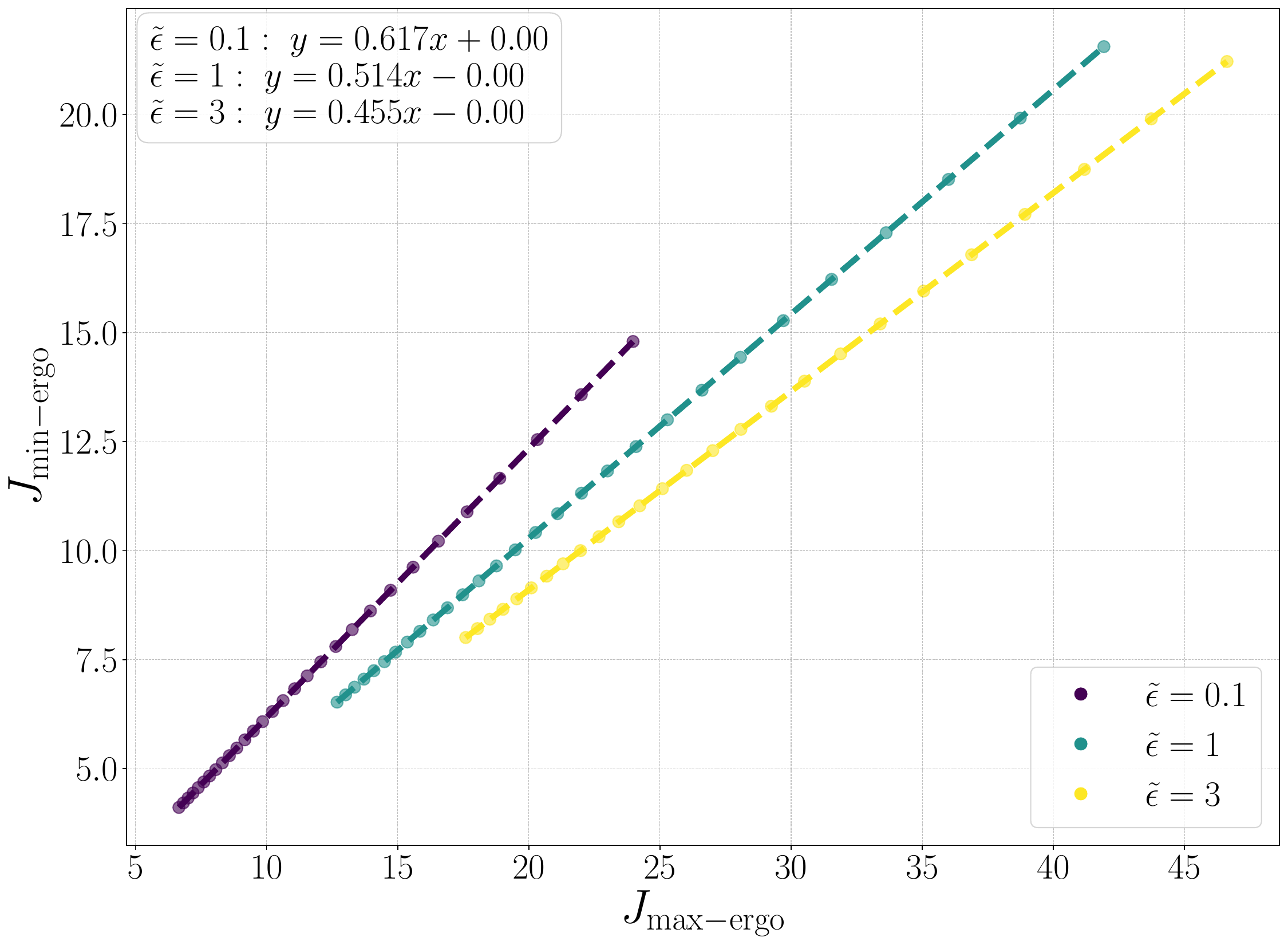}
    \caption{Scaling relation between the angular momentum of the minimum-mass stable ergostar ($J_{\text{min-ergo}}$) and the angular momentum of the maximum-mass stable ergostar ($J_{\text{max-ergo}}$). The legend follows that of the previous figures: points are grouped by color according to their $\tilde{\epsilon}$ value, and the dashed lines represent the corresponding linear fits.} 
    \label{fig: j_j}
\end{figure}

\begin{figure}
    \centering
    \includegraphics[width=0.85\columnwidth]{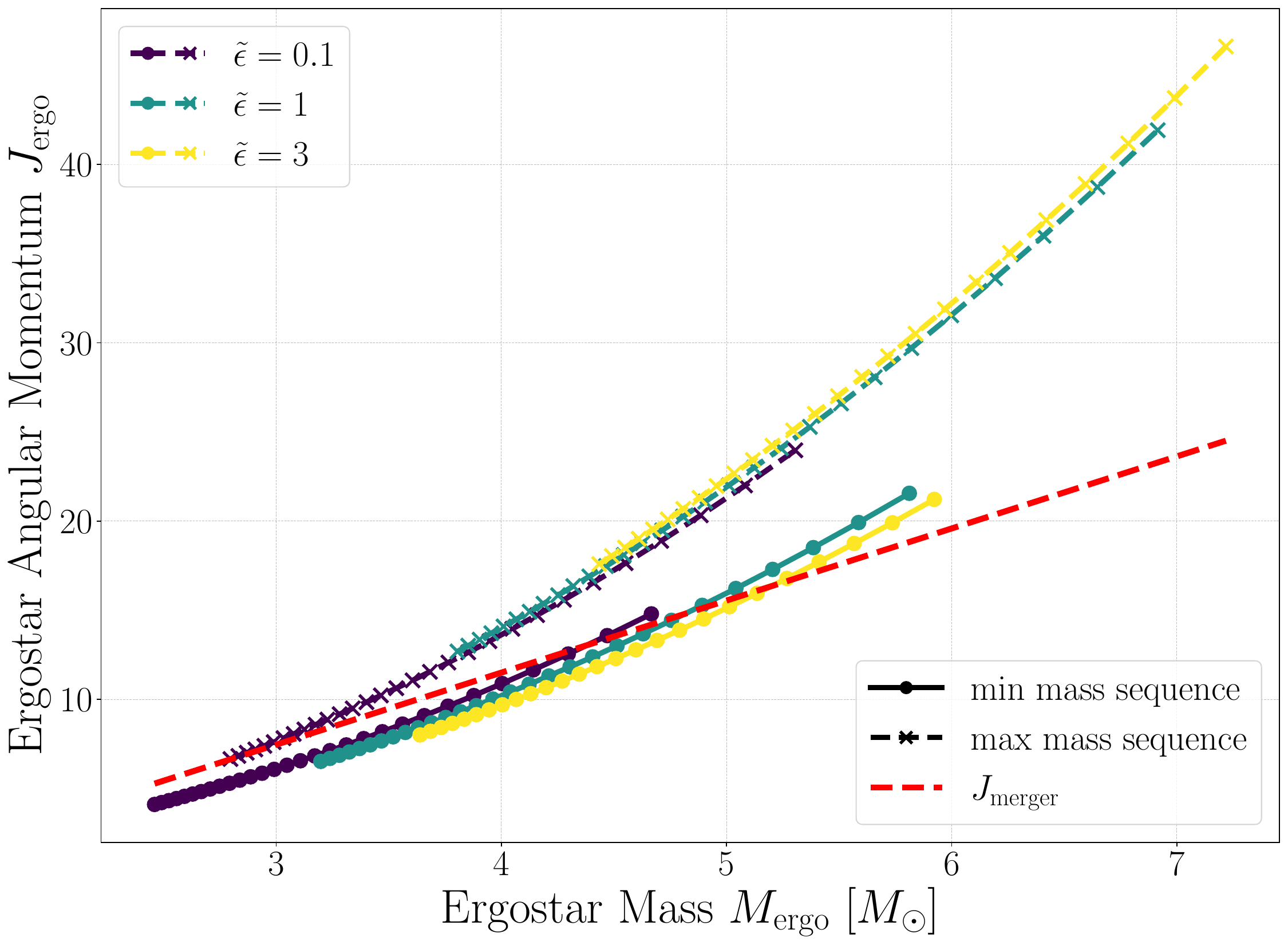}
    \caption{The total angular momentum as functions of the gravitational mass with different parameters. For each value of $\tilde{\epsilon}$, the solid lines with filled circles and the dashed lines with crosses represent the sequences of minimum and maximum mass ergostars, respectively.  The red dashed line represents the empirical threshold for prompt collapse from Bauswein and Stergioulas~\cite{Bauswein_2017}.} 
    \label{fig:j_m}
\end{figure}

Beyond the representative models discussed above, we conducted a systematic parameter survey over the parameter space ($\tilde{\epsilon} \in \{0.1, 1.0, 3.0\} \, \rm MeV$ and $n_{\text{sur}} \in (0.1, 0.36) \, \rm fm^{-3}$) to examine how the EOS parameters affect both ergostar stability and energy extraction. A key feature of the strangeon EOS is the presence of simple scaling relations with respect to the surface baryon density, $n_{\text{sur}}$. 

For a fixed potential depth per quark, $\tilde{\epsilon}$, this manifests as a strong linear correlation between the macroscopic properties (e.g., mass and angular momentum) of stellar models computed for different values of $n_{\text{sur}}$, as explicitly demonstrated in figures~\ref{fig: rho_rho}, \ref{fig: m_m}, and \ref{fig: j_j}. This scaling behavior provides a powerful computational advantage, significantly simplifying the exploration of the parameter space. It allows us to efficiently map the properties across the full range of $n_{\text{sur}}$ and accurately identify the regions of interest for our subsequent analysis on ergostar formation.

Upon examining these scaling relations, a notable distinction emerges. For the central densities (figure~\ref{fig: rho_rho}) and gravitational masses (figure~\ref{fig: m_m}), the linear fits are remarkably similar across the different $\tilde{\epsilon}$ families. The slopes of the best-fit lines exhibit only a very weak dependence on $\tilde{\epsilon}$, resulting in the three dashed lines being nearly coincident. This suggests that the properties of the minimum-mass stable ergostar scale in a nearly universal manner with the properties of the maximum-mass stable ergostar, largely independent of the EOS stiffness.

In stark contrast, the scaling relation for the angular momentum (figure~\ref{fig: j_j}) demonstrates a strong and systematic dependence on $\tilde{\epsilon}$. The slope of the linear fit decreases monotonically as $\tilde{\epsilon}$ increases, with values of $0.62$, $0.51$, and $0.46$ for 
$\tilde{\epsilon}=0.1, 1,$ and $3 \, \rm MeV$, respectively. This trend has a direct physical interpretation that is qualitatively visible in the $J$--$M$ plane (figure~\ref{fig:j_m}). A smaller slope in the $J_{\text{min-ergo}}$--$J_{\text{max-ergo}}$ relation implies a larger vertical separation between the lower and upper boundaries of the stable ergostar domain in the $J$--$M$ plane. Physically, this signifies that a stiffer EOS (larger $\tilde{\epsilon}$) supports a wider range of angular momenta over which stable ergostars can exist.

From our pool of viable candidate models, we selected a subset of twelve representative models to evaluate the maximum extractable energy. The key properties of these models—including the maximum spherical mass, central density, and the maximum extractable energy $\Delta E_{\text{max}}$ along with its allowed variation range—are summarized in table~\ref{tab:eos_properties}. 

Nearly all models we considered can release energy on the order of $0.01 M_\odot$, reinforcing the conclusion that strangeon stars, and in general, compact stars with a sufficiently high density close to their surface\cite{Tsokaros2019, Tsokaros2020b, Tsokaros2020c}, can provide a robust parameter space for postmerger ergostars with energy reservoirs relevant to sGRBs.

\begin{table*}
    \centering
    \caption{Properties of selected cases and the corresponding maximum extractable energy from their ergostar remnants. Columns show the identifier, EOS parameters($\tilde{\epsilon}$ and $n_{sur}$) \cite{Yuan2025}, the maximum  Tolman-Oppenheimer-Volkoff limit ($M_{\text{TOV}}$), its central density ($\rho_{c, \text{TOV}}$), and the calculated maximum extractable energy ($\Delta E_{\text{max}}$) evaluated via $\Delta(M-M_0)$. The table also lists the allowed variation range of the extractable energy, considering a bidirectional baryonic mass fluctuation of $\Delta M_0 = \pm 0.01 M_\odot$ during the post-merger phase.}
    \label{tab:eos_properties}
    \begin{tabular}{|l|cccccc|}
        \hline
        Case ID &$\tilde{\epsilon}$ &$n_{\text{sur}}$ & $M_{\text{TOV}}$  & $\rho_{c, \text{TOV}}$ & $\Delta E_{\text{max}}$ & Range of $\Delta E$\\ 
        [0.1em]
        & $ \rm MeV $ & $ \rm fm^{-3}$ & $M_\odot$ & $\rm g \, cm^{-3}$  & $M_\odot$ & $M_\odot$\\
        \hline
        case A &0.1&0.2061& 2.6877 & $2.23 \times 10^{15}$ & 0.0175 & [0.0163, 0.0193]\\
        case B &0.1&0.2539& 2.5729 &  $2.43 \times 10^{15}$ & 0.0244 & [0.0225, 0.0261]\\
        case C &0.1&0.3069& 2.3400 & $2.94 \times 10^{15}$ & 0.0296 & [0.0278, 0.0312]\\
        case D &0.1&0.3600& 2.1607 & $3.45 \times 10^{15}$ & 0.0196 & [0.0177, 0.0200]\\
        case E &1.0&0.2061& 3.7876 & $1.11 \times 10^{15}$ & 0.0079 & [0.0071, 0.0079]\\
        case F &1.0&0.2539& 3.4128 & $1.37 \times 10^{15}$ & 0.0234 & [0.0206, 0.0253]\\
        case G &1.0&0.3069& 3.1039 & $1.66 \times 10^{15}$ & 0.0307 & [0.0289, 0.0316]\\
        case H &1.0&0.3600& 2.8660 & $1.95 \times 10^{15}$ & 0.0447 & [0.0437, 0.0453]\\
        case I &3.0&0.2061& 4.3558 & $8.89 \times 10^{14}$ & 0.0087 & [0.0087, 0.0118]\\
        case J &3.0&0.2539& 3.9248 & $1.10 \times 10^{15}$ & 0.0283 & [0.0253, 0.0311]\\
        case K &3.0&0.3069& 3.5695 & $1.32 \times 10^{15}$ & 0.0451 & [0.0451, 0.0482]\\
        case L &3.0&0.3600& 3.2960 & $1.55 \times 10^{15}$ & 0.0513 & [0.0513, 0.0558]\\
        \hline
    \end{tabular}
\end{table*}

\section{Discussion}
\label{sec: Discussion}

We have explored the parameter space of dynamically stable ergostars with SnS EOS. In particular, we investigated the possibility of a dynamically stable ergostar remnant that could be formed during BNS mergers and the maximum extractable energy from them. A key finding of this work is that the formation of a dynamically stable, uniformly rotating ergostar as a merger remnant is a robust outcome for a wide range of strangeon matter EOS parameters.
This confirms the results of \cite{Tsokaros2019, Tsokaros2020b, Tsokaros2020c} which have found that the formation of stable ergostars requires a significant density close to the surface of the compact star. Our analysis demonstrates that, for strangeon matter, the post-merger remnant can indeed settle into a transient ergostar state capable of releasing significant energy. 

While the energy extraction potential depends intricately on the geometry of the domain of stable ergostar solutions, a crucial result is that nearly all models we considered can release energy on the order of $0.01 M_\odot$. More importantly, as we have demonstrated through incorporating mass variation bands and testing against strict relativistic causality corrections, this geometrically originated energy budget is exceptionally robust. This represents a substantial and independent energy reservoir, potentially sufficient to power the central engine of a large-scale sGRB or its associated extended emission, irrespective of complex post-merger mass variations such as ejecta or Penrose particle capture. As analyzed in section~\ref{sec:energy}, this robustness is physically guaranteed because such bounded mass fluctuations do not fundamentally alter the stable evolutionary trajectory of the remnant towards the collapse threshold.

It is important to note that demonstrating a sufficient energy budget ($\sim 10^{52}$~erg) is a necessary but not sufficient condition for powering an sGRB. The detailed mechanism of launching a highly relativistic jet must also be considered. For conventional hadronic ergostars, recent GRMHD simulations have revealed a major challenge: magnetic fields tend to strip loosely bound matter from the stellar surface, leading to severe baryon pollution in the polar funnel. This pollution prevents the formation of a force-free environment, resulting in only mildly relativistic outflows ($\Gamma_L \sim 2.5$) rather than sGRB jets \cite{Tsokaros2020c}. However, the situation is physically different for self-bound stars. The surface of self-bound quark matter has been argued to confine baryons while allowing photons, electron--positron pairs, neutrinos, and electromagnetic fields to escape, thereby strongly suppressing baryon pollution originating from the stellar surface \cite{Dai_1998,Paczynski2005,Chen2007}. Since strangeon matter is likewise self-bound by the strong interaction, the same basic mechanism is expected to operate for strangeon ergostars. A complete GRMHD treatment incorporating this surface physics is still required.

The Penrose process is invoked here only as an illustrative example of rotational-energy extraction enabled by negative-energy states, rather than as a specific particle-ejection model. The presence and angular distribution of the ergoregion do not by themselves determine the direction of the resulting outflow. GRMHD simulations indicate that, for both black-hole and long-lived neutron-star merger remnants, jet launching and collimation are mainly controlled by whether the large-scale magnetic-field configuration establishes a magnetically dominated polar funnel or magnetic tower \cite{Ruiz_2018,Combi_2023,Kiuchi_2024,Mbarek_2026}. In the present scenario, the ergoregion provides an additional channel for extracting rotational energy, while the formation and collimation of a possible jet would be governed by the magnetospheric structure. We plan to investigate these processes in future work through dedicated GRMHD simulations with an improved treatment of the finite-density self-bound surface of strangeon matter.

\acknowledgments

This work was supported by the China Postdoctoral Science Foundation (2024M760081), the National Natural Science Foundation of China (12447148), and the National SKA Program of China (2020SKA0120100). E. Z. is supported by NSFC Grant No. 12203017 and the National SKA Program of China No. 2020SKA0120300.


\bibliographystyle{JHEP}
\bibliography{refs}

@PREAMBLE{
 "\providecommand{\noopsort}[1]{}" 
 # "\providecommand{\singleletter}[1]{#1}%" 
}

@article{Yuan2025,
   title={Bayesian inference of strangeon matter using the measurements of PSR J0437-4715 and GW190814},
   volume={111},
   ISSN={2470-0029},
   url={http://dx.doi.org/10.1103/PhysRevD.111.063033},
   DOI={10.1103/physrevd.111.063033},
   number={6},
   journal={Physical Review D},
   publisher={American Physical Society (APS)},
   author={Yuan, Wen-Li and Huang, Chun and Zhang, Chen and Zhou, Enping and Xu, Renxin},
   year={2025},
   month=mar 
}

@article{GW170817,
  title       = {GW170817: Observation of Gravitational Waves from a Binary Neutron Star Inspiral},
  author      = {Abbott, B. P. and others},
  collaboration = {LIGO Scientific Collaboration and Virgo Collaboration},
  journal     = {Phys. Rev. Lett.},
  volume      = {119},
  issue       = {16},
  pages       = {161101},
  year        = {2017},
  month       = {Oct},
  doi         = {10.1103/PhysRevLett.119.161101},
  url         = {https://link.aps.org/doi/10.1103/PhysRevLett.119.161101}
}

@ARTICLE{Friedman1988,
       author = {{Friedman}, John L. and {Ipser}, James R. and {Sorkin}, Rafael D.},
        title = "{Turning Point Method for Axisymmetric Stability of Rotating Relativistic Stars}",
      journal = {\apj},
         year = 1988,
        month = feb,
       volume = {325},
        pages = {722},
          doi = {10.1086/166043},
       adsurl = {https://ui.adsabs.harvard.edu/abs/1988ApJ...325..722F}
}

@article{Blandford1977,
    author = {Blandford, R. D. and Znajek, R. L.},
    title = {Electromagnetic extraction of energy from Kerr black holes},
    journal = {Monthly Notices of the Royal Astronomical Society},
    volume = {179},
    number = {3},
    pages = {433-456},
    year = {1977},
    month = {07},
    issn = {0035-8711},
    doi = {10.1093/mnras/179.3.433},
    url = {https://doi.org/10.1093/mnras/179.3.433},
    eprint = {https://academic.oup.com/mnras/article-pdf/179/3/433/9333653/mnras179-0433.pdf},
}

@article{Komissarov2004,
   title={Electrodynamics of black hole magnetospheres},
   volume={350},
   ISSN={1365-2966},
   url={http://dx.doi.org/10.1111/j.1365-2966.2004.07598.x},
   DOI={10.1111/j.1365-2966.2004.07598.x},
   number={2},
   journal={Monthly Notices of the Royal Astronomical Society},
   publisher={Oxford University Press (OUP)},
   author={Komissarov, S. S.},
   year={2004},
   month=may, pages={427–448} }

@article{Bauswein_2017,
   title={Semi-analytic derivation of the threshold mass for prompt collapse in binary neutron-star mergers},
   volume={471},
   ISSN={1365-2966},
   url={http://dx.doi.org/10.1093/mnras/stx1983},
   DOI={10.1093/mnras/stx1983},
   number={4},
   journal={Monthly Notices of the Royal Astronomical Society},
   publisher={Oxford University Press (OUP)},
   author={Bauswein, Andreas and Stergioulas, Nikolaos},
   year={2017},
   month=aug, pages={4956–4965} 
}

@article{Xu2003,
   title={Solid Quark Stars?},
   volume={596},
   ISSN={1538-4357},
   url={http://dx.doi.org/10.1086/379209},
   DOI={10.1086/379209},
   number={1},
   journal={The Astrophysical Journal},
   publisher={American Astronomical Society},
   author={Xu, R. X.},
   year={2003},
   month=sep, pages={L59–L62} }

@article{Lai2009,
   title={Lennard-Jones quark matter and massive quark stars},
   volume={398},
   ISSN={1745-3925},
   url={http://dx.doi.org/10.1111/j.1745-3933.2009.00701.x},
   DOI={10.1111/j.1745-3933.2009.00701.x},
   number={1},
   journal={Monthly Notices of the Royal Astronomical Society: Letters},
   publisher={Oxford University Press (OUP)},
   author={Lai, X. Y. and Xu, R. X.},
   year={2009},
   month=sep, pages={L31–L35} }

@article{Gao2021,
   title={Rotation and deformation of strangeon stars in the Lennard-Jones model},
   ISSN={1365-2966},
   url={http://dx.doi.org/10.1093/mnras/stab3181},
   DOI={10.1093/mnras/stab3181},
   journal={Monthly Notices of the Royal Astronomical Society},
   publisher={Oxford University Press (OUP)},
   author={Gao, Yong and Lai, Xiao-Yu and Shao, Lijing and Xu, Ren-Xin},
   year={2021},
   month=nov }

@article{Zhang2023_9,
   title={Rescaling strange-cluster stars and its implications on gravitational-wave echoes},
   volume={108},
   ISSN={2470-0029},
   url={http://dx.doi.org/10.1103/PhysRevD.108.063002},
   DOI={10.1103/physrevd.108.063002},
   number={6},
   journal={Physical Review D},
   publisher={American Physical Society (APS)},
   author={Zhang, Chen and Gao, Yong and Xia, Cheng-Jun and Xu, Renxin},
   year={2023},
   month=sep }

@article{Zhang2023_12,
  title = {Hybrid strangeon stars},
  author = {Zhang, Chen and Gao, Yong and Xia, Cheng-Jun and Xu, Renxin},
  journal = {Phys. Rev. D},
  volume = {108},
  issue = {12},
  pages = {123031},
  numpages = {9},
  year = {2023},
  month = {Dec},
  publisher = {American Physical Society},
  doi = {10.1103/PhysRevD.108.123031},
  url = {https://link.aps.org/doi/10.1103/PhysRevD.108.123031}
}

@article{Stergioulas_1995,
   title={Comparing models of rapidly rotating relativistic stars constructed by two numerical methods},
   volume={444},
   ISSN={1538-4357},
   url={http://dx.doi.org/10.1086/175605},
   DOI={10.1086/175605},
   journal={The Astrophysical Journal},
   publisher={American Astronomical Society},
   author={Stergioulas, Nikolaos and Friedman, John L.},
   year={1995},
   month=may, pages={306} }

@article{Tsokaros2020b,
   title={Locating ergostar models in parameter space},
   volume={101},
   ISSN={2470-0029},
   url={http://dx.doi.org/10.1103/PhysRevD.101.064069},
   DOI={10.1103/physrevd.101.064069},
   number={6},
   journal={Physical Review D},
   publisher={American Physical Society (APS)},
   author={Tsokaros, Antonios and Ruiz, Milton and Shapiro, Stuart L.},
   year={2020},
   month=mar }

@article{2017GCN,
  title={GCN CIRCULAR 21507, LIGO/Virgo G298048: INTEGRAL detection of a prompt gamma-ray counterpart},
  author={ Savchenko, V.  and  Mereghetti, S.  and  Ferrigno, C.  and  Kuulkers, E.  and  Bazzano, A.  and  Bozzo, E.  and  Courvoisier, T. J. L.  and Søren Brandt and  Diehl, R.  and  Hanlon, L. },
  year={2017},
}

@article{Savchenko_2017,
   title={INTEGRAL Detection of the First Prompt Gamma-Ray Signal Coincident with the Gravitational-wave Event GW170817},
   volume={848},
   ISSN={2041-8213},
   url={http://dx.doi.org/10.3847/2041-8213/aa8f94},
   DOI={10.3847/2041-8213/aa8f94},
   number={2},
   journal={The Astrophysical Journal Letters},
   publisher={American Astronomical Society},
   author={Savchenko, V. and Ferrigno, C. and Kuulkers, E. and Bazzano, A. and Bozzo, E. and Brandt, S. and Chenevez, J. and Courvoisier, T. J.-L. and Diehl, R. and Domingo, A. and Hanlon, L. and Jourdain, E. and von Kienlin, A. and Laurent, P. and Lebrun, F. and Lutovinov, A. and Martin-Carrillo, A. and Mereghetti, S. and Natalucci, L. and Rodi, J. and Roques, J.-P. and Sunyaev, R. and Ubertini, P.},
   year={2017},
   month=oct, pages={L15} }

@article{Bludman1968,
  title = {Possibility of the Speed of Sound Exceeding the Speed of Light in Ultradense Matter},
  author = {Bludman, S. A. and Ruderman, M. A.},
  journal = {Phys. Rev.},
  volume = {170},
  issue = {5},
  pages = {1176--1184},
  numpages = {0},
  year = {1968},
  month = {Jun},
  publisher = {American Physical Society},
  doi = {10.1103/PhysRev.170.1176},
  url = {https://link.aps.org/doi/10.1103/PhysRev.170.1176}
}

@article{Caporaso1979,
  title = {Must ultrabaric matter be superluminal?},
  author = {Caporaso, G. and Brecher, K.},
  journal = {Phys. Rev. D},
  volume = {20},
  issue = {8},
  pages = {1823--1831},
  numpages = {0},
  year = {1979},
  month = {Oct},
  publisher = {American Physical Society},
  doi = {10.1103/PhysRevD.20.1823},
  url = {https://link.aps.org/doi/10.1103/PhysRevD.20.1823}
}

@ARTICLE{Penrose2002,
       author = {{Penrose}, R.},
        title = "{``Golden Oldie'': Gravitational Collapse: The Role of General Relativity}",
      journal = {General Relativity and Gravitation},
         year = 2002,
        month = jul,
       volume = {7},
        pages = {1141-1165},
          doi = {10.1023/A:1016578408204},
       adsurl = {https://ui.adsabs.harvard.edu/abs/2002GReGr..34.1141P}
}

@article{Komissarov2005,
   title={Observations of the Blandford-Znajek process and the magnetohydrodynamic Penrose process in computer simulations of black hole magnetospheres},
   volume={359},
   ISSN={1365-2966},
   url={http://dx.doi.org/10.1111/j.1365-2966.2005.08974.x},
   DOI={10.1111/j.1365-2966.2005.08974.x},
   number={3},
   journal={Monthly Notices of the Royal Astronomical Society},
   publisher={Oxford University Press (OUP)},
   author={Komissarov, S. S.},
   year={2005},
   month=may, pages={801–808} }

@ARTICLE{2026MNRAS.546ag241L,
       author = {{Li}, Hong-Bo and {Gao}, Shi-Jie and {Li}, Xiang-Dong and {Xu}, Ren-Xin},
        title = "{Polar mounds on strangeon stars: the neutrino emission from ultraluminous X-ray pulsars}",
      journal = {MNRAS},
         year = 2026,
        month = mar,
       volume = {546},
       number = {4},
          eid = {stag241},
        pages = {stag241},
          doi = {10.1093/mnras/stag241},
archivePrefix = {arXiv},
       eprint = {2509.13732},
 primaryClass = {astro-ph.HE},
       adsurl = {https://ui.adsabs.harvard.edu/abs/2026MNRAS.546ag241L}
}

@ARTICLE{2023AdPhX...837433L,
       author = {{Lai}, Xiaoyu and {Xia}, Chengjun and {Xu}, Renxin},
        title = "{Bulk strong matter: the trinity}",
      journal = {Advances in Physics X},
         year = 2023,
        month = jan,
       volume = {8},
       number = {1},
          eid = {2137433},
        pages = {2137433},
          doi = {10.1080/23746149.2022.2137433},
archivePrefix = {arXiv},
       eprint = {2210.01501},
 primaryClass = {hep-ph},
       adsurl = {https://ui.adsabs.harvard.edu/abs/2023AdPhX...837433L}
}

@article{Witten:1984rs,
    author = "Witten, Edward",
    title = "{Cosmic Separation of Phases}",
    reportNumber = "PRINT-84-0400 (IAS,PRINCETON)",
    doi = "10.1103/PhysRevD.30.272",
    journal = "Phys. Rev. D",
    volume = "30",
    pages = "272--285",
    year = "1984"
}

@article{Alcock:1986hz,
    author = "Alcock, Charles and Farhi, Edward and Olinto, Angela",
    title = "{Strange stars}",
    doi = "10.1086/164679",
    journal = "ApJ",
    volume = "310",
    pages = "261--272",
    year = "1986"
}

@inproceedings{Xu:2016uod,
    author = "Xu, Ren Xin and Guo, Yan Jun",
    title = "{Strange Matter: a state before black hole}",
    booktitle = "{11th Rencontres du Vietnam}: {Hot Topics in General Relativity and Gravitation}",
    eprint = "1601.05607",
    archivePrefix = "arXiv",
    primaryClass = "astro-ph.HE",
    doi = "10.1142/9789814699662_0004",
    pages = "119--146",
    year = "2017"
}

@article{Lai:2017ney,
    author = "Lai, Xiao Yu and Xu, Ren Xin",
    title = "{Strangeon and Strangeon Star}",
    eprint = "1701.08463",
    archivePrefix = "arXiv",
    primaryClass = "astro-ph.HE",
    doi = "10.1088/1742-6596/861/1/012027",
    journal = "J. Phys. Conf. Ser.",
    volume = "861",
    number = "1",
    pages = "012027",
    year = "2017"
}

@article{Wang:2024opz,
    author = "Wang, Wei-Yang and Zhang, Chen and Zhou, Enping and Liu, Xiaohui and Niu, Jiarui and Zhou, Zixuan and Gao, He and Liu, Jifeng and Xu, Renxin and Zhang, Bing",
    title = "{On the Energy Budget of Starquake-induced Repeating Fast Radio Bursts}",
    eprint = "2405.07152",
    archivePrefix = "arXiv",
    primaryClass = "astro-ph.HE",
    doi = "10.1088/1674-4527/ad74db",
    journal = "RAA",
    volume = "24",
    number = "10",
    pages = "105012",
    year = "2024"
}

@article{Demorest:2010bx,
    author = "Demorest, Paul and Pennucci, Tim and Ransom, Scott and Roberts, Mallory and Hessels, Jason",
    title = "{Shapiro Delay Measurement of A Two Solar Mass Neutron Star}",
    eprint = "1010.5788",
    archivePrefix = "arXiv",
    primaryClass = "astro-ph.HE",
    doi = "10.1038/nature09466",
    journal = "Nature",
    volume = "467",
    pages = "1081--1083",
    year = "2010"
}

@article{Antoniadis:2013pzd,
    author = "Antoniadis, John and others",
    title = "{A Massive Pulsar in a Compact Relativistic Binary}",
    eprint = "1304.6875",
    archivePrefix = "arXiv",
    primaryClass = "astro-ph.HE",
    doi = "10.1126/science.1233232",
    journal = "Science",
    volume = "340",
    pages = "6131",
    year = "2013"
}

@article{Xu:2003xe,
    author = "Xu, Ren-Xin",
    title = "{Solid quark matter?}",
    eprint = "astro-ph/0302165",
    archivePrefix = "arXiv",
    doi = "10.1086/379209",
     journal = "ApJ",
    volume = "596",
    pages = "L59--L62",
    year = "2003"
}

@ARTICLE{2022MNRAS.516.6172L,
       author = {{Li}, Hong-Bo and {Gao}, Yong and {Shao}, Lijing and {Xu}, Ren-Xin and {Xu}, Rui},
        title = "{Oscillation modes and gravitational waves from strangeon stars}",
      journal = {MNRAS},
         year = 2022,
        month = nov,
       volume = {516},
       number = {4},
        pages = {6172-6179},
          doi = {10.1093/mnras/stac2622},
archivePrefix = {arXiv},
       eprint = {2206.09407},
 primaryClass = {gr-qc},
       adsurl = {https://ui.adsabs.harvard.edu/abs/2022MNRAS.516.6172L}
}

@article{Zhou:2014tba,
    author = "Zhou, E. P. and Lu, J. G. and Tong, H. and Xu, R. X.",
    title = "{Two types of glitches in a solid quark star model}",
    eprint = "1404.2793",
    archivePrefix = "arXiv",
    primaryClass = "astro-ph.HE",
    doi = "10.1093/mnras/stu1370",
    journal = "MNRAS",
    volume = "443",
    number = "3",
    pages = "2705--2710",
    year = "2014"
}

@article{Xu:2006mp,
    author = "Xu, Ren-Xin and Tao, D. J. and Yang, Y.",
    title = "{The superflares of soft Gamma-ray repeatres: Giant quakes in solid quark stars?}",
    eprint = "astro-ph/0607106",
    archivePrefix = "arXiv",
    doi = "10.1111/j.1745-3933.2006.00248.x",
    journal = "MNRAS",
    volume = "373",
    pages = "L85",
    year = "2006"
}

@ARTICLE{2011SCPMA..54.1541D,
       author = {{Dai}, Shi and {Li}, LiXin and {Xu}, RenXin},
        title = "{The plateau of gamma-ray burst: hint for the solidification of quark matter?}",
      journal = {Science China Physics, Mechanics, and Astronomy},
         year = 2011,
        month = aug,
       volume = {54},
       number = {8},
        pages = {1541-1545},
          doi = {10.1007/s11433-011-4384-z},
archivePrefix = {arXiv},
       eprint = {1008.2568},
 primaryClass = {astro-ph.HE},
       adsurl = {https://ui.adsabs.harvard.edu/abs/2011SCPMA..54.1541D}
}

@article{Li:2023tng,
    author = "Li, Hong-Bo and Kang, Yacheng and Hu, Zexin and Shao, Lijing and Xia, Cheng-Jun and Xu, Ren-Xin",
    title = "{Quasi-periodic oscillations during magnetar giant flares in the strangeon star model}",
    eprint = "2309.09847",
    archivePrefix = "arXiv",
    primaryClass = "astro-ph.HE",
    doi = "10.1093/mnras/stad3204",
    journal = "MNRAS",
    volume = "527",
    number = "1",
    pages = "855--862",
    year = "2023"
}

@ARTICLE{2015ApJ...798...56L,
       author = {{Li}, Zhaosheng and {Qu}, Zhijie and {Chen}, Li and {Guo}, Yanjun and {Qu}, Jinlu and {Xu}, Renxin},
        title = "{An Ultra-low-mass and Small-radius Compact Object in 4U 1746-37?}",
      journal = {ApJ},
         year = 2015,
        month = jan,
       volume = {798},
       number = {1},
          eid = {56},
        pages = {56},
          doi = {10.1088/0004-637X/798/1/56},
archivePrefix = {arXiv},
       eprint = {1405.3438},
 primaryClass = {astro-ph.HE},
       adsurl = {https://ui.adsabs.harvard.edu/abs/2015ApJ...798...56L}
}

@article{Wang:2016nqt,
    author = "Wang, Weiyang and Lu, Jiguang and Tong, Hao and Ge, Mingyu and Li, Zhaosheng and Men, Yunpeng and Xu, Renxin",
    title = "{The Optical/UV Excess of X-Ray-dim Isolated Neutron Stars. I. Bremsstrahlung Emission from a Strangeon Star Atmosphere}",
    eprint = "1603.08288",
    archivePrefix = "arXiv",
    primaryClass = "astro-ph.HE",
    doi = "10.3847/1538-4357/aa5e52",
    journal = "ApJ",
    volume = "837",
    number = "1",
    pages = "81",
    year = "2017"
}

@article{Wang:2017hgt,
    author = "Wang, Wei-Yang and Feng, Yi and Lai, Xiao-Yu and Li, Yun-Yang and Lu, Ji-Guang and Chen, Xuelei and Xu, Ren-Xin",
    title = "{The optical/UV excess of X-ray-dim isolated neutron star II. Nonuniformity of plasma on a strangeon star surface}",
    eprint = "1705.03763",
    archivePrefix = "arXiv",
    primaryClass = "astro-ph.HE",
    doi = "10.1088/1674-4527/18/7/82",
    journal = "RAA",
    volume = "18",
    number = "7",
    pages = "082",
    year = "2018"
}

@article{Li:2022qql,
    author = "Li, Hong-Bo and Gao, Yong and Shao, Lijing and Xu, Ren-Xin and Xu, Rui",
    title = "{Oscillation modes and gravitational waves from strangeon stars}",
    eprint = "2206.09407",
    archivePrefix = "arXiv",
    primaryClass = "gr-qc",
    doi = "10.1093/mnras/stac2622",
    journal = "Mon. Not. Roy. Astron. Soc.",
    volume = "516",
    number = "4",
    pages = "6172--6179",
    year = "2022"
}

@article{Zhang_2015,
   title={THE MILLISECOND MAGNETAR CENTRAL ENGINE IN SHORT GRBs},
   volume={805},
   ISSN={1538-4357},
   url={http://dx.doi.org/10.1088/0004-637X/805/2/89},
   DOI={10.1088/0004-637x/805/2/89},
   number={2},
   journal={The Astrophysical Journal},
   publisher={American Astronomical Society},
   author={Lü, Hou-Jun and Zhang, Bing and Lei, Wei-Hua and Li, Ye and Lasky, Paul D},
   year={2015},
   month=may, pages={89} }

@article{Zhang_2001,
   title={Gamma-Ray Burst Afterglow with Continuous Energy Injection: Signature of a Highly Magnetized Millisecond Pulsar},
   volume={552},
   ISSN={0004-637X},
   url={http://dx.doi.org/10.1086/320255},
   DOI={10.1086/320255},
   number={1},
   journal={The Astrophysical Journal},
   publisher={American Astronomical Society},
   author={Zhang, Bing and Mészáros, Peter},
   year={2001},
   month=may, pages={L35–L38} }

@article{Dai_1998,
  title = {$\mathit{\ensuremath{\gamma}}$-Ray Bursts and Afterglows from Rotating Strange Stars and Neutron Stars},
  author = {Dai, Z. G. and Lu, T.},
  journal = {Phys. Rev. Lett.},
  volume = {81},
  issue = {20},
  pages = {4301--4304},
  numpages = {0},
  year = {1998},
  month = {Nov},
  publisher = {American Physical Society},
  doi = {10.1103/PhysRevLett.81.4301},
  url = {https://link.aps.org/doi/10.1103/PhysRevLett.81.4301}
}

@ARTICLE{2025IJMPA..4050180X,
       author = {{Xia}, Chengjun and {Lai}, Xiaoyu and {Xu}, Renxin},
        title = "{Strange matter}",
      journal = {International Journal of Modern Physics A},
         year = 2025,
        month = dec,
       volume = {40},
       number = {34},
          eid = {2550180},
        pages = {2550180},
          doi = {10.1142/S0217751X25501805},
archivePrefix = {arXiv},
       eprint = {2511.01146},
 primaryClass = {astro-ph.HE},
       adsurl = {https://ui.adsabs.harvard.edu/abs/2025IJMPA..4050180X}
}

@ARTICLE{Lu2018,
       author = {{Lu}, JiGuang and {Zhou}, EnPing and {Lai}, XiaoYu and {Xu}, RenXin},
        title = "{Causal propagation of signals in strangeon matter}",
      journal = {Science China Physics, Mechanics, and Astronomy},
         year = 2018,
        month = aug,
       volume = {61},
       number = {8},
          eid = {089511},
        pages = {089511},
          doi = {10.1007/s11433-018-9205-5},
archivePrefix = {arXiv},
       eprint = {1711.08176},
 primaryClass = {astro-ph.HE},
       adsurl = {https://ui.adsabs.harvard.edu/abs/2018SCPMA..6189511L}
}

@ARTICLE{1974AnPhy..86..138B,
       author = {{Bertsch}, G.~F.},
        title = "{Elasticity in the response of nuclei}",
      journal = {Annals of Physics},
         year = 1974,
        month = jul,
       volume = {86},
       number = {1},
        pages = {138-146},
          doi = {10.1016/0003-4916(74)90433-3},
       adsurl = {https://ui.adsabs.harvard.edu/abs/1974AnPhy..86..138B}
}

@ARTICLE{2007PhRvD..76g4026M,
       author = {{Mannarelli}, Massimo and {Rajagopal}, Krishna and {Sharma}, Rishi},
        title = "{Rigidity of crystalline color superconducting quark matter}",
      journal = {\prd},
         year = 2007,
        month = oct,
       volume = {76},
       number = {7},
          eid = {074026},
        pages = {074026},
          doi = {10.1103/PhysRevD.76.074026},
archivePrefix = {arXiv},
       eprint = {hep-ph/0702021},
 primaryClass = {hep-ph},
       adsurl = {https://ui.adsabs.harvard.edu/abs/2007PhRvD..76g4026M}
}

@ARTICLE{KEH1,
       author = {{Komatsu}, Hidemi and {Eriguchi}, Yoshiharu and {Hachisu}, Izumi},
        title = "{Rapidly rotating general relativistic stars. I - Numerical method and its application to uniformly rotating polytropes}",
      journal = {Monthly Notices of the Royal Astronomical Society},
         year = 1989,
        month = mar,
       volume = {237},
        pages = {355-379},
          doi = {10.1093/mnras/237.2.355},
       adsurl = {https://ui.adsabs.harvard.edu/abs/1989MNRAS.237..355K}
}

@article{KEH2,
    author = {Komatsu, Hidemi and Eriguchi, Yoshiharu and Hachisu, Izumi},
    title = {Rapidly rotating general relativistic stars – II. Differentially rotating polytropes},
    journal = {Monthly Notices of the Royal Astronomical Society},
    volume = {239},
    number = {1},
    pages = {153-171},
    year = {1989},
    month = {07},
    issn = {0035-8711},
    doi = {10.1093/mnras/239.1.153},
    url = {https://doi.org/10.1093/mnras/239.1.153},
    eprint = {https://academic.oup.com/mnras/article-pdf/239/1/153/3769884/mnras239-0153.pdf},
}

@ARTICLE{2025Univ...11..354Q,
       author = {{Qi}, Haoyang and {Xu}, Renxin},
        title = "{Strangeon Matter: From Stars to Nuggets}",
      journal = {Universe},
         year = 2025,
        month = oct,
       volume = {11},
       number = {11},
          eid = {354},
        pages = {354},
          doi = {10.3390/universe11110354},
archivePrefix = {arXiv},
       eprint = {2507.13935},
 primaryClass = {astro-ph.HE},
       adsurl = {https://ui.adsabs.harvard.edu/abs/2025Univ...11..354Q}
}

@article{Tsokaros2020a,
  title = {Great Impostors: Extremely Compact, Merging Binary Neutron Stars in the Mass Gap Posing as Binary Black Holes},
  author = {Tsokaros, Antonios and Ruiz, Milton and Shapiro, Stuart L. and Sun, Lunan and Ury\ifmmode \bar{u}\else \={u}\fi{}, K\ifmmode \bar{o}\else \={o}\fi{}ji},
  journal = {Phys. Rev. Lett.},
  volume = {124},
  issue = {7},
  pages = {071101},
  numpages = {6},
  year = {2020},
  month = {Feb},
  publisher = {American Physical Society},
  doi = {10.1103/PhysRevLett.124.071101},
  url = {https://link.aps.org/doi/10.1103/PhysRevLett.124.071101}
}

@article{Tsokaros2020c,
  title = {Magnetic ergostars, jet formation, and gamma-ray bursts: Ergoregions versus horizons},
  author = {Ruiz, Milton and Tsokaros, Antonios and Shapiro, Stuart L. and Nelli, Kyle C. and Qunell, Sam},
  journal = {Phys. Rev. D},
  volume = {102},
  issue = {10},
  pages = {104022},
  numpages = {8},
  year = {2020},
  month = {Nov},
  publisher = {American Physical Society},
  doi = {10.1103/PhysRevD.102.104022},
  url = {https://link.aps.org/doi/10.1103/PhysRevD.102.104022}
}

@article{Tsokaros2019,
  title = {Dynamically Stable Ergostars Exist: General Relativistic Models and Simulations},
  author = {Tsokaros, Antonios and Ruiz, Milton and Sun, Lunan and Shapiro, Stuart L. and Ury\ifmmode \bar{u}\else \={u}\fi{}, K\ifmmode \bar{o}\else \={o}\fi{}ji},
  journal = {Phys. Rev. Lett.},
  volume = {123},
  issue = {23},
  pages = {231103},
  numpages = {5},
  year = {2019},
  month = {Dec},
  publisher = {American Physical Society},
  doi = {10.1103/PhysRevLett.123.231103},
  url = {https://link.aps.org/doi/10.1103/PhysRevLett.123.231103}
}

@ARTICLE{Zhou2024,
       author = {{Zhou}, Yurui and {Zhang}, Chen and {Zhao}, Junjie and {Kiuchi}, Kenta and {Fujibayashi}, Sho and {Zhou}, Enping},
        title = "{Constraints of the maximum mass of quark stars based on postmerger evolutions}",
      journal = {\prd},
         year = 2024,
        month = nov,
       volume = {110},
       number = {10},
          eid = {103012},
        pages = {103012},
          doi = {10.1103/PhysRevD.110.103012},
archivePrefix = {arXiv},
       eprint = {2407.08544},
 primaryClass = {astro-ph.HE},
       adsurl = {https://ui.adsabs.harvard.edu/abs/2024PhRvD.110j3012Z}
}

@article{Zhou2022,
  title = {Evolution of equal mass binary bare quark stars in full general relativity: Could a supramassive merger remnant experience prompt collapse?},
  author = {Zhou, Enping and Kiuchi, Kenta and Shibata, Masaru and Tsokaros, Antonios and Ury\ifmmode \bar{u}\else \={u}\fi{}, K\ifmmode \bar{o}\else \={o}\fi{}ji},
  journal = {Phys. Rev. D},
  volume = {106},
  issue = {10},
  pages = {103030},
  numpages = {7},
  year = {2022},
  month = {Nov},
  publisher = {American Physical Society},
  doi = {10.1103/PhysRevD.106.103030},
  url = {https://link.aps.org/doi/10.1103/PhysRevD.106.103030}
}

@article{Ruiz_2018,
   title={GW170817, general relativistic magnetohydrodynamic simulations, and the neutron star maximum mass},
   volume={97},
   ISSN={2470-0029},
   url={http://dx.doi.org/10.1103/PhysRevD.97.021501},
   DOI={10.1103/physrevd.97.021501},
   number={2},
   journal={Physical Review D},
   publisher={American Physical Society (APS)},
   author={Ruiz, Milton and Shapiro, Stuart L. and Tsokaros, Antonios},
   year={2018},
   month=Jan }

@misc{Kiuchi_2024,
      title={A large-scale magnetic field produced by a solar-like dynamo in binary neutron star mergers}, 
      author={Kenta Kiuchi and Alexis Reboul-Salze and Masaru Shibata and Yuichiro Sekiguchi},
      year={2024},
      eprint={2306.15721},
      archivePrefix={arXiv},
      primaryClass={astro-ph.HE},
      url={https://arxiv.org/abs/2306.15721}, 
}

@misc{Combi_2023,
      title={Jets from neutron-star merger remnants and massive blue kilonovae}, 
      author={Luciano Combi and Daniel M. Siegel},
      year={2023},
      eprint={2303.12284},
      archivePrefix={arXiv},
      primaryClass={astro-ph.HE},
      url={https://arxiv.org/abs/2303.12284}, 
}

@article{Mbarek_2026,
   title={Hadronic Processes, Plasma Evolution, and Neutrino Emission in Magnetic Towers of Neutron Star Merger Remnants},
   volume={1005},
   ISSN={2041-8213},
   url={http://dx.doi.org/10.3847/2041-8213/ae7971},
   DOI={10.3847/2041-8213/ae7971},
   number={1},
   journal={The Astrophysical Journal Letters},
   publisher={American Astronomical Society},
   author={Mbarek, Rostom and Wu, Jiaxi and Most, Elias R.},
   year={2026},
   month=June, pages={L12} }

@article{Paczynski2005,
   title={Gamma-ray bursts from quark stars},
   volume={362},
   ISSN={1745-3925},
   url={http://dx.doi.org/10.1111/j.1745-3933.2005.00059.x},
   DOI={10.1111/j.1745-3933.2005.00059.x},
   number={1},
   journal={Monthly Notices of the Royal Astronomical Society: Letters},
   publisher={Oxford University Press (OUP)},
   author={Paczyński, B. and Haensel, P.},
   year={2005},
   month=Sept, pages={L4–L7} }

@article{Chen2007,
   title={The Birth of Quark Stars: Photon-driven Supernovae?},
   volume={668},
   ISSN={1538-4357},
   url={http://dx.doi.org/10.1086/522777},
   DOI={10.1086/522777},
   number={1},
   journal={The Astrophysical Journal},
   publisher={American Astronomical Society},
   author={Chen, Anbo and Yu, Tianhong and Xu, Renxin},
   year={2007},
   month=Sept, pages={L55–L58} }

\end{document}